\documentstyle[twocolumn,aps,epsfig,graphicx]{revtex}
\newcommand{\be}{\begin{equation}}
\newcommand{\ee}{\end{equation}}
\newcommand{\bea}{\begin{eqnarray}}
\newcommand{\eea}{\end{eqnarray}}
\newcommand{\psat}{p_{\rm sat}}
\renewcommand{\d}{{{\rm d}}}

\newcommand{\lton}{\mathrel{\lower.9ex
                  \hbox{$\stackrel{\displaystyle <}{\sim}$}}}

\begin{document}
\title{The Effective Pressure of a Saturated Gluon Plasma}
\author{Adrian Dumitru and Miklos Gyulassy}
\address{Physics Department, Columbia University,
538W 120th Street, New York, NY 10027, USA}
\date{\today}
\maketitle   
\begin{abstract}
The evolution of the gluon plasma produced with 
saturation initial conditions is calculated
via transport theory for nuclear collisions with $0.1<\surd{s}<10 A$~TeV.
The effective longitudinal pressure is found to remain
significantly below the lattice QCD pressure with these initial
conditions until the plasma cools to near the confinement scale.
The absolute value of the transverse energy per unit of
rapidity and its dependence on beam energy
is shown to provide a sensitive test of gluon saturation models since
the fractional transverse energy loss due to final state interactions
is predicted to be much smaller and exhibit a weaker energy dependence
than nondissipative hydrodynamics applied throughout the evolution.
\end{abstract}
\narrowtext

There is an ongoing experimental
program to produce a (transient) deconfined phase
of QCD matter~\cite{lattice}  in the laboratory via nuclear
collisions at high energies~\cite{qm99}. 
The  production mechanism is the liberation
of a large number of gluons from the nuclear structure
functions.  
The plasma is produced from copious minijet gluons at central rapidity,
 $y\simeq0$,  with
transverse momentum $p_T>p_0$. The rapidity density of gluons
liberated in central $A+A$ collisions can be estimated
from~\cite{Collins:1985ue}
\bea
\frac{\d N}{\d y}(p_0) &=& K \;T_{AA}(b=0)
\int\limits_{p_T>p_0} \d^2p_T
\int \d x_a \d x_b\, \nonumber\\
& & G\left(x_a,p_T^2\right) 
G\left(x_b,p_T^2\right) \frac{\hat{s}}{\pi}
  \frac{\d\sigma}{\d \hat{t}} 
  \delta\left(\hat{s}+\hat{t}+
  \hat{u}\right)\quad.
\label{minijet}
\eea
$\hat{s}$, $\hat{t}$, $\hat{u}$ are the Mandelstam variables 
for the
parton-parton scattering process, and $\d\sigma/\d\hat{t}$
denotes the hard-scattering differential cross section in lowest order of
perturbative QCD. $G(x,p_T^2)$ denotes the LO gluon  distribution function
in the nucleus. The phenomenological factor $K=2$ accounts
approximately for NLO corrections. The nuclear overlap function
$T_{AA}(0)={A^2}/{\pi R_A^2}$ determines the number of binary nucleon
collisions in head-on reactions within the Glauber approach, where
$R_A\simeq1.1A^{1/3}$~fm for mass $A$ nuclei.

For large $p_0$, the produced gluon plasma is dilute. As
$p_0$ decreases, however, the density of gluons
increases rapidly due to the increase
of $G(x,p_T^2)$ as $x\approx 2p_T/\surd s$ decreases. It has been
conjectured~\cite{glr} 
 that below  some transverse
momentum scale $p_0\le \psat$ the phase-space density of produced gluons
may saturate since $gg\rightarrow g$
recombination could limit further growth of the structure functions.
Phenomenologically, this condition may arise when
gluons (per unit rapidity
and  transverse area $\pi/\psat^2$) become closely packed and
 fill the available nuclear interaction transverse area.
The saturation scale $\psat$ can  thus be estimated from
\be \label{sat_std}
\frac{\d N}{\d y}(\psat)=\psat^2 R_A^2/\beta \quad,
\ee
where $\beta\sim 1$.
For $\beta=1$ the solution reported  in EKRT~\cite{kimmo}
was 
\bea
\psat &\approx& 0.208A^{\; 0.128}\,\sqrt{s}^{\; 0.191} \nonumber\\
C_1 
%\equiv \d E_T\d y/\psat\d N\d y
 &\approx& 1.34A^{-0.007}\,\sqrt{s}^{0.021}~, \label{kimmo_param}
\eea
where $\psat$ and $\sqrt{s}$ are in units of GeV and $C_1$ is the
average transverse
energy per gluon (in units of $\psat$). The focus of this paper
is to investigate  whether the final observed $\d E_T^f/\d y$ can be used
to test the predicted $A$ and $\surd s$ dependence of
  the  initial  $\d E_T^i/\d y=C_1\psat \d N(\psat)/\d y$.

Different gluon saturation models 
based on non-linear evolution and classical Yang-Mills
equations~\cite{McLerran:1994ka,Mueller:2000fp} suggest that
the factor $\beta$ in (\ref{sat_std}) may vary parametrically as
\be\label{Gexpr}
\beta(\psat) = \frac{4\pi\alpha(\psat)N_c}{c (N_c^2-1)}\quad,
\ee
where $c\sim 1$ is a nonperturbative factor proportional to the
fraction of the initial gluons in the nucleus which are liberated.
This factor was recently estimated  using lattice classical Yang-Mills
methods~\cite{KrVe} to be $c\approx 1.3$. 

The first data~\cite{phobos} from RHIC on Au+Au collisions
 at $\surd{s}= 130A$~GeV
with $\d N_{ch}/\d\eta\approx 560$ is in fact 
reproduced by the EKRT saturation model
~\cite{kimmo} with $\beta = 1$ assuming isentropic expansion
(see \cite{wg00}). 
On the other hand,
a fit to the Phobos data using eq.(\ref{Gexpr})
requires $c\approx 1.9$. Note that
the solution of eq.~(\ref{sat_std}) with $\beta\ne 1$ can be obtained from
eq.~(\ref{kimmo_param}) by rescaling the mass-number
$A\rightarrow A/\beta^{2/3}$, and iterating until the stationary point is
reached. Extrapolating to CERN-LHC energy
 the minijet multiplicities are predicted to be $\d N/\d y=3200$ for
$\beta=1$ versus $\d N/\d y=5100$ for $\beta(\psat)$ using eq.~(\ref{Gexpr}).

While $\d N(A,\surd s)/\d y$ systematics  
provide one experimental handle 
to test different  saturation and fixed scale models
of initial conditions\cite{wg00}, another important observable 
that probes collective dynamics is
the transverse energy per unit rapidity, $\d E_T/\d y$.
In EKRT the final value of $\d E_T^{f}/\d y$
was predicted to be  much smaller than produced initially
due to collective longitudinal work  assuming the
validity of isentropic hydrodynamics.

If the expansion proceeds in approximate local equilibrium with
pressure $p= c^2 \epsilon$ and speed of sound $c$, then 
the energy density, $\epsilon(\tau)$, must decrease faster than the expansion
rate $\Gamma_{exp}=1/\tau$ and
leads to a  bulk  transverse energy loss
\be \label{edens_ansatz}
\frac{E_T(\tau)}{E_T(\tau_0)} =\frac{\tau\epsilon}{\tau_0\epsilon_0}= 
\left(\frac{\tau_0}{\tau}
\right)^\delta\quad.
\ee
{\em If} local equilibrium
is maintained during the evolution $\delta=c^2$. In contrast, if the system
expands too rapidly to maintain local equilibrium, then
the effective pressure is reduced (relative  to that from LQCD)
due to dissipation.
The extreme asymptotically free plasma case corresponds
to free streaming with
$\delta=0$. $E_T$ thus provides an important
 barometric observable
that  probes the (longitudinal) pressure in the  plasma~\cite{Gyulassy:1984ub}.
There have been of course many studies on the magnitude of dissipative
effects on this and other observables, see
e.g.~\cite{Hoskaj,Danielewicz:1985ww,zpc,Wong,HsW}.
The new twist on this old problem that  we consider  here
is to extend those studies to the novel initial
conditions suggested by gluon saturation 
models~\cite{kimmo,Mueller:2000fp,KrVe}.

To compute the transverse energy loss due to longitudinal work, we 
 employ the
Boltzmann equation in relaxation time approximation~\cite{Hoskaj,HsW,Baym},
\be \label{BE_relax}
p\cdot\partial f(p,x)=\Gamma_{rel} \, p\cdot u \left(
f_{eq}(p\cdot u)-f(p,x)\right)~.
\ee
$u^\mu$ denotes the four-velocity of the comoving frame and $f_{eq}$ is
the chemical and thermal equilibrium phase space distribution, 
towards which $f$ evolves at a relaxation rate $\Gamma_{rel}$.
It is important to emphasize that this much simplified transport equation
has been extensively tested against full 3+1D covariant parton cascade
codes~\cite{zpc} and provides a surprisingly accurate equation for calculating
the evolution of  the transverse energy observable
even  in highly dissipative
systems far from equilibrium ($\Gamma_{rel}\lton \Gamma_{exp}$). 

The relevant relaxation rate is given by the fractional 
energy loss per unit length,
\be
\Gamma_{rel} = \frac{1}{E}\frac{\d E}{\d z}~,
\ee
which receives a contribution both from elastic and inelastic scattering,
\bea
\Gamma_{rel} &=& \rho \int \left(\d\sigma_{el} \frac{\Delta E_{el}}{E} 
+ \d\sigma_{in} \frac{\Delta E_{in}}{E}\right) \nonumber\\
&\approx& \rho \int_{\mu^2}^{Q^2} \d q^2 \frac{\d\sigma_{el}}{\d q^2}
\left\{ \frac{q^2}{2E^2}+
\frac{\alpha N_c}{\pi} 
\int_{\mu^2}^{q^2} \frac{\d k_T^2}{k_T^2} \int\frac{\d x}{x}
x \right\} \nonumber\\
&=& \rho\left(\frac{16\pi\alpha^2 N_c^2}{N_c^2-1}\right)
\left(\frac{1}{s}\log\frac{Q^2}{\mu^2} +\frac{\alpha N_c}{\pi\mu^2}
\log\frac{\overline{q^2}}{\mu^2}\right).
\eea
In these equations $\rho(\tau)$ denotes the gluon density in the
local restframe, $\mu^2$ is the Debye screening mass in the medium,
$Q^2\simeq s$ is the upper bound for the momentum transfer in the scattering
process, and $x$ denotes the fraction of energy carried away  by 
radiated gluons.
In the last step we replaced the momentum transfer $q^2$ in the expression for
the radiative energy loss by its average,
%\be
$\overline{q^2}
%= \int \d\sigma_{el} \,\,\, q^2 \Big/
%\int \d\sigma_{el} = 
\approx \mu^2\log{Q^2}/{\mu^2}.$
%~.\ee
In local thermal equilibrium the average energy per gluon and the Debye
screening scale are
\bea
\overline{s}/2 &=& \left({\epsilon}/{\rho}\right)^2 \simeq 9T^2\\
\mu^2 &=& N_c g^2 T^2/3 = 4\pi\alpha T^2.
\eea
Assuming that the ratio $\overline{s}/\mu^2$ is essentially the same
even out of equilibrium, the
relaxation rate is approximately given by
\be \label{relax_rate}
\Gamma_{rel} \approx 9\pi\alpha^2 \frac{\rho^3}{\epsilon^2}
\left(\log\frac{1}{\alpha} + \frac{27}{2\pi^2}\right)
\equiv K_{in} 9\pi\alpha^2 \frac{\rho^3}{\epsilon^2}
\log\frac{1}{\alpha}~.
\ee
We have set the double-logarithm of $Q^2/\mu^2\sim1/\alpha$ equal to unity.
The expression~(\ref{relax_rate}) in fact overestimates the relaxation
rate at early times, because the screening length $1/\mu\sim
1/g\psat$ exceeds formally the horizon at $\tau_0=1/\psat$ for
longitudinal Bjorken expansion~\cite{Bj} and because 
we neglect the suppression
of radiation due to formation time physics. However, since with
$\psat \lton 2$ GeV,   $g\approx 2$  up
to the LHC energy domain,
we ignore this formal point in the  discussion below.

The inelastic, radiative energy loss represents a significant source of
uncertainty and is especially important in chemically undersaturated
models of the initial conditions~\cite{Wong,Geiger}.
Within the saturation model, gluon multiplication through $2\rightarrow3$
processes may lead to thermalization of the soft radiated
gluons at times parametrically large as compared to $\tau_0$, while
the effect on the hard part of the gluon distribution is 
small~\cite{mueller_son}. In the present paper we do not attempt  a
more detailed treatment of radiative energy loss but simply
 vary the factor $K_{in}\sim 1-2$ to provide a measure
of the theoretical uncertainites. 
We note that radiative energy loss for gluons with modest
 $p_T < 5$  GeV,  is in any case significantly suppressed
due to finite kinematic constraints and destructive
interference effects~\cite{glv}.  

The EKRT saturation model
predicts the gluon density to be nearly chemically saturated 
already at the initial
time $\tau_0=1/\psat$. This  follows from the observation 
that ideal-gas formulas
$\rho\sim T^3$, $\epsilon\sim T^4$ applied with chemical potential
$\mu_g=0$ yield the same
``temperature'' $T_0$~\cite{kimmo}.

At the initial time $\tau_0=1/\psat$, $\overline{s}/2=\epsilon_0^2/
\rho_0^2=C_1^2\psat^2$.
Therefore, noting that the comoving gluon density at
time $\tau_0$ is $\rho_0=\psat^3/\pi \beta$, the ratio of
the relaxation  rate to the expansion rate is given by
\be \label{scatt_to_exp}
\frac{\Gamma_{rel}}{\Gamma_{exp}}=
 K_{in} \frac{9\alpha^2}{\beta C_1^2}
 \log\frac{1}{\alpha}\quad,
\ee
While $\Gamma_{rel}\propto \psat$ increases  as a power of the energy
in eq.~(\ref{kimmo_param}), the Bjorken boundary conditions~\cite{Bj}
force the system to expand londitudinally initially also at 
an increasing rate $\Gamma_{exp}(\tau_0)=\psat$. 
The essential quantity that fixes the magnitude of the effective
pressure relative to that predicted by LQCD
is the ratio of rates in Eq.~(\ref{scatt_to_exp}), which dimensionally
is simply a function of $\alpha(\psat)$.
The asymptotic freedom property~\cite{Gross:1973id} of QCD therefore
requires that this ratio vanishes as $\surd{s}\rightarrow \infty$.
In  (\ref{scatt_to_exp}) the rate of how fast it vanishes is controlled by 
$\alpha^2 K_{in}(\psat)/\beta(\psat)$. Therefore, with
saturation initial conditions, asymptotic freedom reduces the {\em effective}
pressure acting at early times
$\tau\sim \tau_0$ and causes the initial evolution to deviate
from ideal hydrodynamics for a time  interval  that, as we show
below, increases with energy.

The total number of interactions during the evolution up to time $\tau$
is given by
\be \label{phifunc}
\phi(\tau) \equiv \int_{\tau_0}^\tau\d\tau' \Gamma_{rel}(\tau')
\simeq \frac{K_{in}}{\beta} \frac{9\alpha^2\log(1/
\alpha)}{C_1^2}\log{\frac{\tau}{\tau_0}} \;\; .
\ee
 The last expression holds
close to the free streaming regime (``Knudsen limit''), where
the number of scatterings increases only logarithmically with time.
%Also for this estimate we neglected the time dependence of
%$\alpha$ for simplicity. 
%For both saturation models 
We find that $\phi$ reaches on the order of unity at $\tau^*\approx 0.4$~fm 
in the BNL-RHIC to CERN-LHC energy region.
However, the local relaxation rate at $\tau^*$
is still less than or on the order of the expansion rate ($1/\tau^*$).
In  a {\em non-expanding} plasma
$\phi\sim 1$ provides a rough equilibration criterion.
However, as long as $\Gamma_{rel}$ is not significantly larger than
$\Gamma_{exp}$ this criterion is insufficient
to address how much  collective 
hydrodynamic work can be done  by the plasma.

For  a quantitative estimate, we  must solve the  kinetic equations
(\ref{BE_relax}). 
The first energy moment of that equation together with  energy
conservation to replace $\epsilon_{eq}$ by $\epsilon(\tau)$,
results in the energy density evolution equation ~\cite{Hoskaj,HsW,Baym}
\be \label{edensrelax}
e^{\phi}\frac{\tau\epsilon}{\tau_0\epsilon_0}=1+\int_0^\phi
\d\phi' e^{\phi'}\frac{\tau'(\phi')\epsilon(\phi')}{\tau_0\epsilon_0}
h\left(\frac{\tau'(\phi')}{\tau(\phi)}\right)~.
\ee
This equation applies if the  initially produced
partons have a vanishing longitudinal momentum spread in the comoving frame,
i.e.\ assuming a strong correlation between space-time rapidity and
momentum space rapidity~\cite{Bj}.
The function $h(x)$ appearing in~(\ref{edensrelax}) is given by
$2 h(x)\sqrt{1-x^2}=x\sqrt{1-x^2}+\arcsin\sqrt{1-x^2}$;
it insures that $\delta\rightarrow1/3$ as $\tau\rightarrow \infty$.

Expanding~(\ref{edensrelax}) to first order in $\phi$ yields
\be
\frac{\tau\epsilon}{\tau_0\epsilon_0}=1-
\left(\frac{3}{4}-\frac{\pi^2}{16}\right) \phi + {\cal O}\left(\phi^2\right)~,
\ee
giving for $\tau\sim \tau_0$
\be \label{analyt}
\delta = \left(\frac{3}{4}-\frac{\pi^2}{16}\right) 
\frac{9K_{in} \alpha^2}
{\beta C_1^2}\log\frac{1}{\alpha}~.
\ee
Note that $\delta\rightarrow 0$ 
as $\psat\rightarrow\infty$ in accordance with the discussion above.

For our numerical estimates  we let  $\alpha$  creep with time.
Since the effective temperature scales as $\epsilon(\tau)^{1/4}$,
the effective coupling, $\alpha(T_{eff})$,
increases slowly with time approximately as
\be
\alpha(\tau)= \left(\frac{12 \pi}{27}\right) 
/\log\left(1+\frac{\psat^2}{\Lambda_{QCD}^2}
\left(\frac{\epsilon(\tau)}{\epsilon(\tau_0)}\right)^{1/2} \right)
\;\; , \label{als}
\ee
with $\Lambda_{QCD}=200$ MeV. Furthermore,  longitudinal expansion
 constrains $\rho(\tau)\tau$ to remain constant. 
As the system cools, the mean center of mass energy in collisions also
decreases as 
\be
\bar{s}(\tau)=2\left(\frac{\epsilon(\tau)}{\rho(\tau)}\right)^2=
\bar{s}(\tau_0)\left(\frac{\tau \epsilon(\tau)}{\tau_0 \epsilon(\tau_0)}\right)^2~.
\ee
The number of collisions between $\tau_0$ and $\tau$
in this case is given by
\bea
\phi(\tau) &=& \Gamma_{rel}(\tau_0) \int_{\tau_0}^\tau 
{d\tau^\prime}
\left(\frac{\alpha(\tau^\prime)}{\alpha(\tau_0)}\right)^2 
\left(\frac{\epsilon(\tau_0)}{\epsilon(\tau^\prime)}\right)^{2}  
\left(\frac{\tau_0}{\tau^\prime}\right)^3 \nonumber\\
& &\times \frac{\log (1+1/\alpha(\tau^\prime))}{\log( 1+1/\alpha(\tau_0))}~.
\label{kernel}
\eea
We regulated the logarithmic dependence above for numerical stability.
Note that with eq.~(\ref{kernel}), eq.~(\ref{edensrelax}) 
is a nonlinear self-consistency equation for $\epsilon(\tau)$. 
\begin{figure}[htp]
\centerline{\hbox{\epsfig{figure=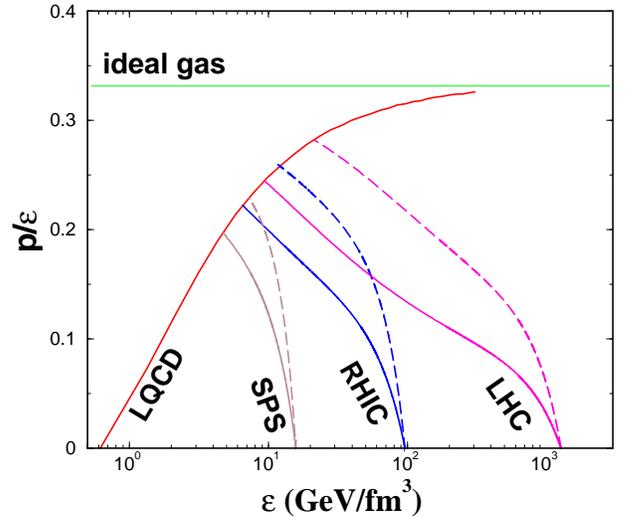,width=8cm}}}
\caption{
The ratio of effective longitudinal pressure to energy density
as a function of the energy density along the dynamical path is shown
for SPS, RHIC, and LHC saturation initial conditions\protect\cite{kimmo}.
Solid (dashed) curves are for $\beta=1$ and $K_{in}=1(2)$.
The LQCD equation of state~\protect\cite{lattice}
is also shown for comparison.}
\label{poe}
\end{figure}
In Figure~\ref{poe} we show the effective
longitudinal pressure as a function of the
energy density for $\surd s=20, 200, 5400$ A GeV saturation
 initial conditions. The ratio $p/\epsilon$
is  defined here by $\delta(\tau)$ as obtained solving
eq.~(\ref{edensrelax}) numerically.
Our definition of the effective pressure absorbes {\em all} disspative
corrections
to the perfect fluid equation, $u_\mu \partial_\nu [(
\epsilon+p)u^\mu u^\nu-pg^{\mu\nu}]=\d\epsilon/\d\tau+(\epsilon+p)/\tau=0$.
For comparison the pressure of equilibrium QCD is also shown
for  $N_f=3$. This curve is obtained from the
$N_f=0$ lattice data of~\cite{lattice} rescaling the 
number of relativistic degrees of freedom by 47.5/16, and assuming a
transition temperature $T_c(N_f=3)=160$~MeV.

Initially $p/\epsilon$ starts at zero in this model
and remains small for a large time relative to $1/\psat$ 
because the plasma is torn apart by the initial
rapid longitudinal expansion.
The effective pressure 
approaches the LQCD curve from below and reaches it at a time
$\tau_L\approx 1-2$~fm at RHIC $\surd {s}=200$~AGeV, by which
time the energy density 
has dropped by an order of magnitude, $\epsilon_L\equiv\epsilon(\tau_L)
=6.5-12$~GeV/fm$^3$.
For LHC $\surd {s}=5400$~AGeV,  $\tau_L=3-7$~fm during which
the energy density falls by almost two orders of magnitude to $\epsilon_L=
9.5-21.5$~GeV/fm$^3$. The quoted intervals correspond to $K_{in}=1-2$
using the EKRT parametrization~(\ref{kimmo_param}).
The contrast between the dynamical path followed by the saturated plasma
compared to the equilibrium equation of state is striking.
A qualitatively similar behavior of the early
logitudinal pressure has also been found from solutions of
diffusion equations~\cite{Jeff}.
\begin{figure}[htp]
\centerline{\hbox{\epsfig{figure=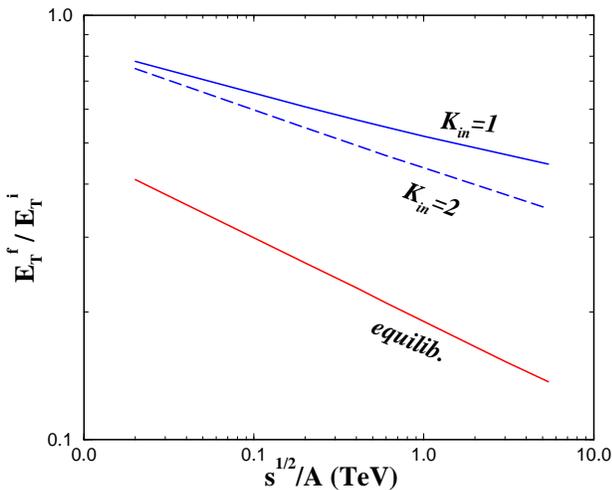,width=8cm}}}
\caption{The ratios of the final to the initial
transverse energy per unit rapidity
are shown as a function of beam energy for central
Au+Au collisions. The initial value corresponds to the EKRT
parametrization ($\beta=1$) and the transport results are for $K_{in}=1,2$.
For comparison, the final transverse energy
assuming that local equilibrium was maintained throughout the evolution
is also shown.}
\label{et}
\end{figure}
Since the longitudinal gradients at $\tau_L$ are much smaller
than at $\tau_0$, the evolution beyond $\tau_L$
is much more likely to  follow isentropic
hydrodynamics along the lattice QCD equation
of state. In this case one could calculate detailed
differential hadronic observables
along the same lines as in ~\cite{DumRi}
using the conditions at $\tau_L$ as the initial conditions
for  3+1D hydrodynamics.

The main experimentally
 observable consequence of the reduced effective pressure
is  shown in Fig.~\ref{et}. The ratio ${E_T^f}/{E_T^i} = 
{\tau_f\epsilon_f}/{\tau_0
\epsilon_0}$ has been obtained from the solution of the transport equation
assuming $\epsilon_f=2$~GeV/fm$^3$ which corresponds
roughly to $T\simeq T_c$. $\tau_f$ is estimated
assuming hydrodynamic expansion from the point were the
trajectories in Fig.~\ref{poe} reach the LQCD curve at time $\tau_L$, with the
equation of state ${p}/{\epsilon}=a+b\log\epsilon$ ($\epsilon$ in units
of GeV/fm$^3$). The parameters $a=0.051$, $b=0.092$ provide a reasonable
fit to the LQCD curve shown in Fig.~\ref{poe}. In this case,
\be
\frac{\tau_f}{\tau_L}= \left(\frac{1+a+b\log\epsilon_L}
{1+a+b\log\epsilon_f}\right)^{1/b}~.
\ee

On the other hand, if ideal hydrodynamics were applicable already at
$\tau_0$, the final observed transverse energy for 1+1 dimensional
adiabatic expansion would be
\be \label{adiab_exp}
\frac{E_T^f}{E_T^i} = \frac{\tau_f\epsilon_f}{\tau_0\epsilon_0}
= \frac{\tau_f(T_f s_f - p_f)}{\tau_0(T_0 s_0 - p_0)} = \frac{T_f}{T_0}.
\ee
The last step follows both for $p_{0,f}\simeq0$ as well as
$p_{0,f}= T_{0,f} s_{0,f}/4$ from the condition of entropy conservation,
$\tau s={\rm const}$~\cite{kimmo}.
Strong transverse expansion leads to slightly larger
$E_T^f$ but we shall neglect that small effect here for simplicity.
Clearly, for $T_f\sim T_c\approx160$~MeV one would observe a much smaller
transverse energy in the final state than in the initial state.
%predicted
%by the saturation model. 
Moreover, $E_T^f/E_T^i$ would also have
significantly stronger energy dependence such that $E_T^f$ deviates
more and more from $E_T^i$ with increasing $\surd s$.
In this sense isentropic hydrodynamics
erases information on the interesting initial conditions via  this observable.
The solutions of the transport equations clearly show a smaller decrease of
$E_T^f$ and of the logarithmic slope, $\kappa=\d \log E_T^f/\d\log \surd s$,
due to final state interactions. We find that  $\kappa=0.50$ for the
initial state~(\ref{kimmo_param}) evolved 
with $K_{in}=1$, $\kappa=0.46$ with  $K_{in}=2$, while 
$\kappa=0.40$ with isentropic  expansion, eq.~(\ref{adiab_exp}).
For comparison, the initial EKRT saturated
$E_T^i=\pi R_A^2 \tau_0\epsilon_0$ scales with the higher power
$\kappa=0.59$ according to eq.~(\ref{kimmo_param}).
The fractional transverse energy loss is thus  less
dependent on energy than
for entropy conserving expansion for which
$E_T^f/E_T^i\propto 1/T_0\propto
1/\sqrt {s}^{\; 0.2}$.
This is due to the increasingly long time spent far from equilibrium in 
Fig.~\ref{poe} as the beam energy increases.

The results in Fig.~\ref{et} are 
encouraging from the point of view of searching for evidence 
of gluon saturation 
in nuclei at high energies.
Experimental data on $\d E_T/\d y$ or $\d E_T/\d \eta$
for central Au+Au collisions at RHIC will soon provide a new  test of
saturation and non-saturation models at those energies. Since
we predict that dissipative effects reduce considerably
the effective longitudinal 
pressure in Fig.~1, 
the beam energy dependence of the transverse energy
is expected to reflect much more accurately 
the predicted power law dependence of the initial
conditions as seen in Fig.~2.
We therefore conclude that  the energy and $A$ systematics of
the bulk calorimetric observable,
$\d E_T/\d y$, will be a sensitive test of saturation models
of gluon plasmas produced in the RHIC to LHC energy range.

\acknowledgements
We thank K.\ Eskola,
K.\ Kajantie, L.\ McLerran, A.H.\ Mueller, D.\ Son, R.\ Venugopalan
and K.\ Tuominen for
helpful criticism and discussions on thermalization aspects and saturation.
We thank the BNL nuclear theory group for 
hosting a stimulating workshop  during which this work was  completed.
We acknowledge support from the DOE Research Grant, 
Contract No.DE-FG-02-93ER-40764. 

\end{document}